\documentclass[aps,prb,twocolumn,showpacs,floatfix,superscriptaddress] {revtex4}
\usepackage[T1]{fontenc}
\usepackage[latin9]{inputenc}

\usepackage{graphicx}
\usepackage{amssymb}
\usepackage{amsmath}
\usepackage{esint}
\usepackage{color}
\usepackage{soul}



\begin{document}

\newcommand{\be}{\begin{equation}}
\newcommand{\ee}{  \end{equation}}
\newcommand{\ba}{\begin{eqnarray}}
\newcommand{\ea}{  \end{eqnarray}}
\newcommand{\ve}{\varepsilon}

\title{Spin transport and tunable Gilbert damping in a single-molecule magnet junction}

\author{Milena Filipovi\'{c}} 
\affiliation{Fachbereich Physik, Universit\"at Konstanz, D-78457 Konstanz, Germany}

\author{Cecilia Holmqvist}
\affiliation{Fachbereich Physik, Universit\"at Konstanz, D-78457 Konstanz, Germany}

\author{Federica Haupt} 
\affiliation{Institut f\"ur Theorie der Statistischen Physik, RWTH Aachen, D-52056 Aachen, Germany }

\author{Wolfgang Belzig}
\affiliation{Fachbereich Physik, Universit\"at Konstanz, D-78457 Konstanz, Germany}

\date{\today}
\begin{abstract}
We study time-dependent electronic and spin transport through an electronic level connected to
two leads and coupled with a single-molecule magnet via exchange 
interaction. The molecular spin is treated as a classical 
variable and precesses around an external magnetic field. 
We derive expressions for charge and spin currents by 
means of the Keldysh nonequilibrium Green's functions technique in linear order
with respect to the time-dependent magnetic field created by this precession. 
The coupling between the electronic spins and the magnetization dynamics of the molecule 
creates inelastic tunneling processes which contribute to the spin currents. The inelastic 
spin currents, in turn, generate a spin-transfer torque acting 
on the molecular spin. This back-action includes a contribution to 
the Gilbert damping and a modification of the precession frequency.
The Gilbert damping coefficient can be controlled by the bias and gate 
voltages or via the external magnetic field and has a nonmonotonic dependence 
on the tunneling rates.
\end{abstract}
\maketitle

\section{Introduction}

Single-molecule magnets (SMMs) are quantum magnets, i.e., mesoscopic quantum objects 
with a permanent magnetization. They are typically formed by paramagnetic ions stabilized 
by surrounding organic ligands.\cite{Christou} 
SMMs show both classical properties such as magnetization hysteresis \cite{hysteresis} and quantum properties such as 
spin tunneling,\cite{spintunneling} coherence,\cite{coherence} and quantum phase interference.\cite{hysteresis,interference}
They have recently been in the center of interest \cite{hysteresis,ses,Gatteschi}
in view of their possible applications as information storage\cite{Ventra} and processing devices. \cite{Flat}

Currently, a goal in the field of nanophysics is to control and manipulate individual quantum systems, in particular, individual 
spins.\cite{Awschalom,Bog} 
Some theoretical works have investigated electronic transport 
through a molecular magnet contacted to leads.\cite{Kimm,Leu,Romeike,Cecil,Bode,Shnirman,Mosshammer,CTimm} In this case, the transport properties are modified due to the 
exchange interaction between the itinerant electrons and the SMM, \cite{Timm} making it possible to read out the spin state of the molecule using transport currents.
Conversely, the spin dynamics and hence the state of an SMM can also be controlled by transport currents.
Efficient control of the molecule's spin state can be achieved by coupling to ferromagnetic contacts as well. \cite{Elste}

Experiments have addressed the electronic transport properties through magnetic molecules such as Mn$_{12}$ and Fe$_{8}$, \cite{Heersche,Jo} 
which have been intensively studied as they are promising candidates for memory devices.\cite{LeuenbergerLoss}
Various phenomena such as large conductance gaps,\cite{Voss1} switching behavior,\cite{Choi} negative differential conductance,
dependence of the transport on magnetic fields and Coulomb blockades have been experimentally observed. \cite{Heersche,Jo,Zyazin,Roch}
Experimental techniques, including, for instance, scanning tunneling microscopy (STM),\cite{Heersche,Jo,Hirjibehedin,Zhou,Kahle} break 
junctions,\cite{Reed} and three-terminal 
devices,\cite{Heersche,Jo,Zyazin} have
been employed to measure electronic transport through an SMM.
Scanning tunneling spectroscopy and STM experiments show that quantum properties of 
SMMs are preserved when deposited on substrates.\cite{Kahle}
The Kondo effect in SMMs with magnetic anisotropy has been investigated both theoretically\cite{Romeike} and experimentally.\cite{Otte,Parks}
It has been suggested\cite{Delgado} and experimentally verified\cite{Loth} that a spin-polarized tip can be used to control the magnetic state
of a single Mn atom.

In some limits, the large spin $S$ of an SMM can be treated as a 
classical magnetic moment.
In that case, the spin dynamics is described by the Landau-Lifshitz-Gilbert (LLG) equation that incorporates effects of external 
magnetic fields as well as torques originating from damping phenomena. \cite{LandLif,Gilbert}
In tunnel junctions with magnetic particles, LLG equations have been derived using
perturbative couplings\cite{Fransson,Kiesslich} and the nonequilibrium Born-Oppenheimer approximation.\cite{Bode} Current-induced magnetization switching is driven by a generated spin-transfer torque (STT)\cite{slonczewski,berger,Tserkovnyak,Ralph} as a back-action effect of the electronic spin transport on the magnetic particle.\cite{Slon,Sankey,Bode,Chudnovskiy}
A spin-polarized STM (Ref. 36) has been used to experimentally study STTs in relation to a molecular magnetization.\cite{Krause}
This experimental achievement opens new possibilities for data storage technology and applications using current-induced STTs.

The goal of this paper is to study the interplay between electronic spin currents and the spin dynamics of an SMM.
We focus on the spin-transport properties of a tunnel junction through which transport occurs via a single electronic energy level in the presence of an SMM.
The electronic level may belong to a neighboring quantum dot (QD) or it may be an orbital related to the SMM itself. The electronic level and the 
molecular spin are coupled via exchange interaction, allowing for interaction between the spins of the itinerant electrons tunneling through 
the electronic level and the spin dynamics of the SMM.
We use a semiclassical approach in which
the magnetization of the SMM is treated as a classical spin whose dynamics is controlled by an external magnetic field,
while for the electronic spin and charge transport we use instead a quantum description.
The magnetic field is assumed to be constant, leading to a precessional motion of the spin around the magnetic field axis.
The electronic level is subjected both to the effects of the molecular spin and the external magnetic field,  generating a Zeeman split of 
the level. The spin precession makes additional channels available for transport, which leads to the possibility of precession-assisted inelastic tunneling. During a tunnel event, spin-angular
momentum may be transferred between the inelastic spin currents and the molecular spin,
leading to an STT that may be used to manipulate the spin of the SMM. 
This torque includes the so-called {\it Gilbert damping}, which is a phenomenologically introduced damping term of the LLG
equation, \cite{Gilbert} and a term corresponding to a modification of the precession frequency.
We show that the STT and hence the SMM's spin dynamics can be controlled by the external magnetic field, the bias voltage across the junction, 
and the gate voltage acting on the electronic level.

The paper is organized as follows:
We introduce our model and formalism
based on the Keldysh nonequilibrium Green's functions technique \cite{Jauho1993,Jauho1994, JauhoBook} in Sec. \ref{sec: model}, where we derive 
expressions for the charge and spin currents in linear order with 
respect to a time-dependent magnetic field and analyze the spin-transport properties at zero temperature.
In Sec. III we replace the general magnetic field of Sec. \ref{sec: model} by an SMM whose spin precesses in an external constant magnetic field, calculate the 
STT components related to the Gilbert damping, and the modification of the precession frequency, and analyze 
the effects of the external magnetic field as well as the bias and gate voltages on
the spin dynamics.
Conclusions are given in Sec. IV.

\section{Current response to a time dependent magnetic field}\label{sec: model}
\subsection{Model and Formalism}
\begin{figure}[t]
\includegraphics[width=8.2cm,keepaspectratio=true]{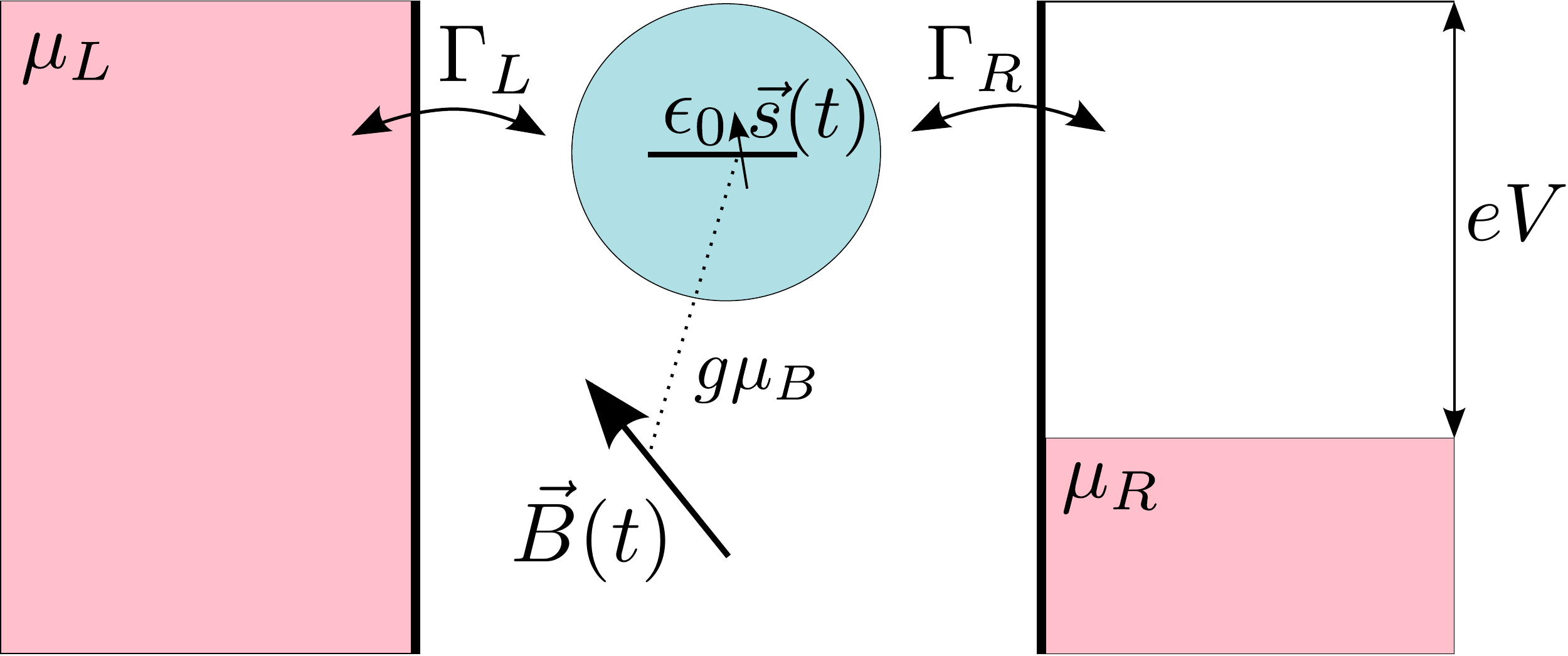}
\caption{\label{fig: qd and b field}(Color online) A quantum dot with a single electronic level $\epsilon_{0}$ coupled to two metallic 
leads with chemical potentials $\mu_{L}$ and $\mu_{R}$ in the presence of an external time-dependent magnetic field ${\vec B}(t)$. 
The spin-transport properties of the junction are determined by the bias voltage $eV=\mu_L-\mu_R$, the position of the 
level $\epsilon_0$, the tunnel rates $\Gamma_{L}$ and $\Gamma_{R}$, and the magnetic field.}
\end{figure}

For the sake of clarity, we start by considering a junction consisting of a noninteracting single-level QD coupled with two normal, metallic 
leads in the presence of an external, time-dependent magnetic field (see Fig.~\ref{fig: qd and b field}).
The leads are assumed to be noninteracting and unaffected by the external field.
The total Hamiltonian describing the junction is given by $\hat{H}(t)=\hat{H}_{L,R}+\hat{H}_{T}+\hat{H}_D(t)$. The Hamiltonian of the free electrons in the leads reads
$\hat{H}_{L,R}=\sum_{k,\sigma,\xi\in \{L,R\}}\epsilon_{k\sigma\xi} \hat{c}^\dagger_{k\sigma\xi} \hat{c}_{k\sigma\xi}$, where $\xi$ denotes the 
left ($L$) or right ($R$) lead,
whereas the tunnel coupling between the QD
and the leads can be written as $\hat{H}_{T}=\sum_{{k,\sigma,\xi\in L,R}}  [V_{k\xi}
\hat{c}^\dagger_{k\sigma\xi} \hat{d}_{\sigma}+V^{\ast}_{k\xi}
\hat{d}^\dagger_{\sigma} \hat{c}_{k\sigma\xi}]$. The spin-independent tunnel matrix element is given by $V_{k\xi}$. 
The 
operators $ \hat{c}^\dagger_{k\sigma\xi}(\hat{c}_{k\sigma\xi})$ and $ \hat{d}^\dagger_{\sigma} (\hat{d}_{\sigma})$ are 
the creation (annihilation) operators of the electrons in the leads and the QD, respectively.
The subscript $\sigma=\uparrow,\downarrow$ denotes the spin-up or spin-down 
state of the electrons.
The electronic level $\epsilon_{0}$ of the QD is influenced by an external magnetic field $\vec{B}(t)$ consisting of 
a constant part $\vec{B}^c$ and a time-dependent part $\vec{B}'(t)$.
The Hamiltonian of the QD describing the interaction between the electronic spin $\hat{\vec{s}}$ and the magnetic field
is then given by $\hat{H}_D(t)=\hat{H}^c_D+\hat{H}'(t)$, where the constant and time-dependent parts 
are $\hat{H}^c_D=\sum_{{\sigma}}\epsilon_{0} \hat{d}^\dagger_{\sigma} \hat{d}_{\sigma}+g\mu_{B}\hat{\vec{s}}\vec B^c$ and $\hat{H}'(t)=g\mu_{B}\hat{\vec{s}}\vec{B}'(t)$.
The proportionality factor $g$ is the gyromagnetic ratio of the electron and $\mu_{B}$ is the Bohr magneton.

The average charge and spin currents from the left lead to the electronic level are given by 
\begin{equation}
I_{L\nu}(t) =q_\nu\bigg \langle \frac{d}{dt} \hat{N}_{L\nu} \bigg \rangle\ =  q_\nu\frac{i}{\hbar} 
\big \langle \big [\hat{H},\hat{N}_{L\nu} \big ] \big \rangle,
\end{equation}
where $\hat{N}_{L\nu}=\sum_{{k,\sigma,\sigma\prime}}
\hat{c}^\dagger_{k\sigma L}(\hat{\sigma}_\nu)_{\sigma\sigma^\prime} 
\hat{c}_{k\sigma\prime L}$ is the charge and spin occupation number operator of the left contact. The index $\nu=0$ corresponds to the charge 
current, while $\nu= x,y,z$ indicates the different components of the spin-polarized current. The current coefficients $q_\nu$ 
are then $q_0=-e$ and $q_{\nu\neq 0}=\hbar/2$. In addition, it is useful to define the vector ${\hat\sigma_\nu}=(\hat 1,\hat{\vec\sigma})$, 
where $\hat 1$ is the identity operator and $\hat{\vec\sigma}$ consists of the Pauli operators with matrix elements $(\hat{\vec{\sigma}})_{\sigma \sigma^\prime}$.
Using the Keldysh nonequilibrium Green's functions technique, the currents can then be obtained as \cite{Jauho1994, JauhoBook}
\begin{align}
I_{L\nu}{(t)} =&-\frac{2q_\nu}{\hbar}{\rm Re}\int dt^\prime {\rm Tr}\big\{\hat\sigma_\nu[ {\hat{G}}^{r} {(t,t^\prime)}{\hat{\Sigma}}^{<}_{L}(t^\prime,t)\\
& \quad \quad \quad\quad \quad\quad +{\hat{G}}^{<} {(t,t^\prime)}{\hat{\Sigma}}^{a}_{L}(t^\prime,t) ]\big\},\nonumber
\end{align}
where $\hat{G}^{r,a,<}$ are the retarded, advanced, and lesser Green's functions of the electrons
in the QD with the matrix elements $G^{r,a}_{\sigma\sigma^\prime} {(t,t^\prime)}=\mp i\theta(\pm t \mp t^\prime)\langle\{\hat{d}_{\sigma}{(t)},
\hat{d}^\dagger_{\sigma^\prime} {(t^\prime)}\}\rangle$ and $G^<_{\sigma\sigma^\prime} (t,t^\prime)= i \langle \hat{d}^\dagger_{\sigma^\prime} (t^\prime) \hat{d}_\sigma(t)\rangle$, while $\hat{\Sigma}^{r,a,<}_{L}(t,t^\prime)$ 
are self-energies from the coupling between the QD
and the left lead. Their nonzero matrix elements are diagonal in the electronic spin space with respect to the basis of eigenstates of $\hat{s_{z}}$, given by $\Sigma^{r,a,<}_{L}(t,t')=\sum_{{k}}
V_{kL}g^{r,a,<}_{kL}{(t,t')}V^{\ast}_{kL}$.
The Green's functions $g^{r,a,<}_{k L}{(t,t')}$ are the retarded, advanced and lesser Green's functions of the free electrons in the left lead.
The retarded Green's functions $\hat{G}^{r}_{0}$ of the electrons in the QD, in
the presence of the constant magnetic field $\vec B^c$, are found using the equation of motion technique,\cite{Bruus} while the lesser Green's functions 
$\hat{G}^{<}_{0}$ are obtained from the Keldysh equation ${\hat{G}}^{<}_{0}={\hat{G}}^{r}_{0}{\hat{\Sigma}}^{<}{\hat{G}}^{a}_{0}$, where multiplication implies internal time integrations.\cite{JauhoBook} 
The time-dependent
part of the magnetic field can be expressed as $\vec{B}'(t)= \sum_{\omega} 
(\vec B_{\omega}e^{-i\omega t}+\vec B_{\omega}^{\ast}e^{i\omega t})$, where $\vec B_{\omega}$ 
is a complex amplitude.
This magnetic field acts as a time-dependent perturbation that
can be expressed as $\hat{H}^\prime(t)=\sum_{\omega} ({\hat{H}}_{\omega} e^{-i\omega t}+{\hat{H}}^
{\dagger}_{\omega} e^{i\omega t})$, where ${\hat{H}}_{\omega}$ is an operator in the electronic spin space and its matrix representaton in the 
basis of eigenstates of ${\hat{s}}_z$ is given by
\begin{equation}
{\hat{H}}_{\omega}=\frac{g\mu_{B}}{2}\left (
   \begin{array}{cc}
   B_{\omega z} & B_{\omega x}-iB_{\omega y} \\
   B_{\omega x}+iB_{\omega y} & -B_{\omega z}
   \end{array}
   \right ).
\end{equation}
Applying Dyson's expansion, analytic continuation rules and the Keldysh equation,\cite{JauhoBook} one obtains a first-order approximation of the Green's functions
describing the electrons in the QD that can be written as
\begin{eqnarray}
{\hat{G}}^{r} &\approx& {\hat{G}}^{r}_{0}+{\hat{G}}^{r}_{0}\hat{H}^\prime{\hat{G}}_{0}^{r}, \\ \nonumber
{\hat{G}}^{<} &\approx& {\hat{G}}^{r}_{0}{\hat{\Sigma}}^{<}{\hat{G}}^{a}_{0}+{\hat{G}}^{r}_{0}\hat{H}^\prime
{\hat{G}}_{0}^{r}{\hat{\Sigma}}^{<}{\hat{G}}_{0}^{a}+ {\hat{G}}^{r}_{0}{\hat{\Sigma}}^{<}{\hat{G}}^{a}_{0}
\hat{H}^\prime{\hat{G}}^{a}_{0}.
\end{eqnarray}
The expression for the currents in this linear approximation is given by
\begin{align}\label{eq: current linear response}
I_{L\nu}{(t)}&=-\frac{2q_\nu}{\hbar}{\rm Re}\,{\rm  Tr}\big\{{\hat\sigma_\nu}[{\hat{G}}_{0}^{r}{\hat{\Sigma}}^{<}_{L}+
{\hat{G}}_{0}^{<} {\hat{\Sigma}}^{a}_{L}\\
& +{\hat{G}}_{0}^{r}\hat{H}^\prime{\hat{G}}_{0}^{r} {\hat{\Sigma}}^{<}_{L}+{\hat{G}}_{0}^{r}\hat{H}^\prime{\hat{G}}_{0}^{<} 
{\hat{\Sigma}}^{a}_{L}+{\hat{G}}_{0}^{<}\hat{H}^\prime{\hat{G}}_{0}^{a} {\hat{\Sigma}}^{a}_{L}]\big\}.\nonumber
\end{align}
Eq. (\ref{eq: current linear response}) is then Fourier transformed
in the wide-band limit, in which the level width function,
$\Gamma(\epsilon)=-2\,{\rm Im} \{ \Sigma^{r}(\epsilon) \}$, is 
constant, ${\rm Re} \{ \Sigma^{r}(\epsilon) \}=0$, and
one can hence write the retarded self-energy originating from the dot-lead coupling as
$\Sigma^{r,a}(\epsilon)=\mp i\Gamma/2$.
From this transformation, one obtains
\begin{equation}\label{eq: J of t}
 I_{L\nu}{(t)}= I^{{\rm dc}}_{L\nu}+ \sum_{\omega}[I_{L\nu}(\omega)e^{-i\omega t}+I_{L\nu}^{\ast}(\omega)e^{i\omega t}].
\end{equation}
Using units in which $\hbar=1$, the dc part of the currents \cite{JauhoBook}  $I^{{\rm dc}}_{L\nu}$ and the time-independent complex 
components $I_{L\nu}(\omega)$ are given by
\begin{equation}
 I^{\rm dc}_{L\nu}={q_\nu}\int\frac{d\epsilon}{\pi} \frac{\Gamma_{L}\Gamma_{R}}{\Gamma} \ 
[f_{L}(\epsilon)-f_{R}(\epsilon)]\,{\rm Tr}\, {\rm Im}\{{\hat\sigma_\nu{\hat{G}}_{0}^{r}(\epsilon)}\}
\end{equation}
and
\begin{align}\label{eq: spin current}
&I_{L\nu}(\omega)=-i{q_\nu}\int\frac{d\epsilon}{2\pi} \frac{\Gamma_{L}\Gamma_{R}}{\Gamma}\Big\{[f_{L}(\epsilon)-f_{R}(\epsilon)]\\
&\times{\rm Tr}\{\hat\sigma_\nu[{\hat{G}}_{0}^{r}(\epsilon+\omega){\hat{H}}_{\omega}{\hat{G}}_{0}^{r}(\epsilon)+
2i\, {\rm Im}\{{\hat{G}}_{0}^{r}(\epsilon)\}{\hat{H}}_{\omega}{\hat{G}}_{0}^{a}(\epsilon-\omega)]\}\nonumber\\
&+\sum_{\xi=L,R}\frac{\Gamma_{\xi}}{\Gamma_{R}}[f_\xi(\epsilon-\omega)-f_{L}(\epsilon)]
\, {\rm Tr}[\hat\sigma_\nu{\hat{G}}_{0}^{r}(\epsilon){\hat{H}}_{\omega}{\hat{G}}_{0}^{a}(\epsilon-\omega)]\Big\}.\nonumber
\end{align}
In the above expressions, $f_{\xi}(\epsilon)=[e^{(\epsilon-\mu_{\xi})/k_{B}T}+1]^{-1}$ is the Fermi distribution of the electrons in lead $\xi$,
where $k_{B}$ is the Boltzmann constant. The retarded Green's function $\hat{G}_{0}^{r}(\epsilon)$ is given by 
$\hat{G}_{0}^{r}(\epsilon)=[\epsilon-\epsilon_{0}-\Sigma^{r}(\epsilon)-(1/2)g\mu_{B}\hat{\vec{\sigma}} \vec B^{c}]^{-1}$.\cite{Bode}

The linear response of the spin current with respect to the applied time-dependent magnetic field can be expressed in terms of complex spin-current 
susceptibilities defined as
\begin{eqnarray}
{\chi}^{L}_{\nu j}(\omega)=\frac{\partial I_{L\nu}(\omega)}{\partial B_{\omega j}}, \quad  j=x,y,z.
\end{eqnarray}
The complex components $I_{L\nu}(\omega)$ are conversely given by $I_{L\nu}(\omega)=\sum_{j}\chi^L_{\nu j}(\omega)B_{\omega j}$.
By taking into account that $\partial{\hat{H}}_{\omega}/\partial B_{\omega j}=(1/2)g\mu_{B}\hat\sigma_{j}$ and using Eq. (\ref{eq: spin current}),
the current susceptibilities can be written as
\begin{align}
&\chi^{L}_{\nu j}(\omega)=-iq_\nu g\mu_{B}\int\frac{d\epsilon}{4\pi} \frac{\Gamma_{L}\Gamma_{R}}{\Gamma}\Big\{[f_{L}(\epsilon)-f_{R}(\epsilon)]\\
&\times{\rm Tr}\{\hat\sigma_\nu[{\hat{G}}_{0}^{r}(\epsilon+\omega)\hat\sigma_{j}{\hat{G}}_{0}^{r}(\epsilon)+
2i\, {\rm Im}\{{\hat{G}}_{0}^{r}(\epsilon)\}{\hat\sigma_{j}}{\hat{G}}_{0}^{a}(\epsilon-\omega)]\}\nonumber\\
&+\sum_{\xi}\frac{\Gamma_{\xi}}{\Gamma_{R}}[f_{\xi}(\epsilon-\omega)-f_{L}(\epsilon)]
{\rm Tr}[\hat\sigma_\nu{\hat{G}}_{0}^{r}(\epsilon){\hat\sigma_j}{\hat{G}}_{0}^{a}(\epsilon-\omega)]\Big\}.\nonumber
\end{align}
The components obey ${\chi}^{L}_{\nu j}(-\omega)={\chi}^{L\ast}_{\nu j}(\omega)$.
In other words, they satisfy the Kramers-Kronig relations\cite{LandauBook} that can be written in a compact form as
\begin{eqnarray}\label{eq: Kramers-Kronig}
{\chi}^{L}_{\nu j}(\omega)=\frac{1}{i\pi}P\int_{-\infty}^{\infty}\frac{\chi^{L}_{\nu j}(\xi)}{\xi-\omega}d\xi,
\end{eqnarray}
with $P$ denoting the principal value.

For any $i,j,k=x,y,z$ such that $j\neq k$ and $j,k\neq i$, where $i$ indicates the direction of the constant part of the
magnetic field $\vec B^c=B^c\vec{e}_{i}$, the complex current
susceptibilities satisfy the relations
\begin{align}
 {\chi}^{L}_{jj}(\omega)&={\chi}^{L}_{kk}(\omega)  \label{eq: susc rel 1} \\
{\rm and} \quad \quad {\chi}^{L}_{jk}(\omega)&=-{\chi}^{L}_{kj}(\omega),   \label{eq: susc rel 2}  \
\end{align}
in addition to Eq. (\ref{eq: Kramers-Kronig}).
The other nonzero components are $\chi^{L}_{0i}(\omega)$ and $\chi^{L}_{ii}(\omega)$.
In the absence of a constant magnetic field, the only nonvanishing components obey ${\chi}^{L}_{xx}(\omega)={\chi}^{L}_{yy}(\omega)={\chi}^{L}_{zz}(\omega)$.

Finally, the average value of the electronic spin in the QD reads
$\vec{s}(t)=\langle\hat{\vec{s}}(t)\rangle =(1/2) \sum_{\sigma \sigma^\prime} \vec{\sigma}_{\sigma \sigma^\prime} \langle\hat{d}^\dagger_\sigma(t) \hat{d}_{\sigma^\prime}(t)\rangle =-(i/2) \sum_{\sigma \sigma^\prime} 
\vec{\sigma}_{\sigma \sigma^\prime}{\hat G}^{<}_{\sigma'\sigma}(t,t)$ and the complex spin susceptibilities are defined as
\begin{eqnarray}
{\chi}^{s}_{i j}(\omega)=\frac{\partial s_{i}(\omega)}{\partial B_{\omega j}}.
\end{eqnarray}
They represent the linear responses of the electronic spin components 
to the applied time-dependent magnetic field and satisfy the relations similar to Eqs. (\ref{eq: Kramers-Kronig}), (\ref{eq: susc rel 1}), and (\ref{eq: susc rel 2}) 
given above.

\subsection{Analysis of the spin and current responses}

\begin{figure}[t]
\includegraphics[height=5.1cm,keepaspectratio=true]{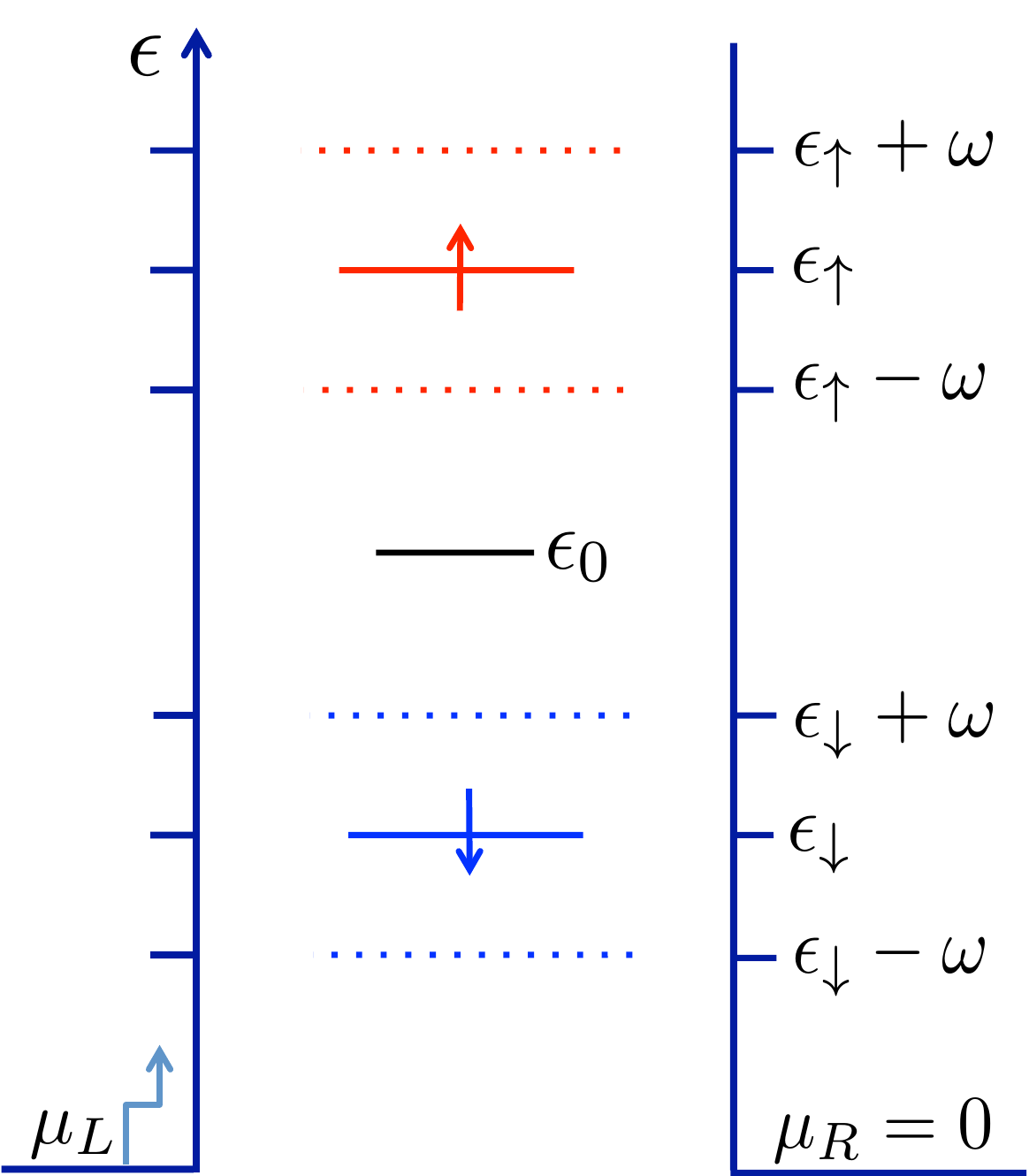}
\caption{(Color online) Sketch of the electronic energy levels of the QD in the presence of a time-dependent magnetic field. In a static magnetic field, the electronic level $\epsilon_0$ (solid black line) 
splits into
the Zeeman levels $\epsilon_{\uparrow,\downarrow}$ (solid red and blue lines). If the magnetic field in addition to the static component 
also includes a time-dependent part with a characteristic frequency $\omega$, additional levels appear at energies 
$\epsilon_{\uparrow}\pm\omega$ (dotted red lines) and $\epsilon_{\downarrow}\pm\omega$ (dotted blue lines). Hence, there are 
six channels available for transport.}\label{fig: energy levels}
\end{figure}

We start by analyzing the transport properties of the junction at zero temperature in response to the external time-dependent 
magnetic field $\vec B(t)$. The constant component of the magnetic field $\vec B^c$ 
generates a Zeeman split of the QD level $\epsilon_0$, resulting in the levels $\epsilon_{\uparrow,\downarrow}$, 
where $\epsilon_{\uparrow,\downarrow}=\epsilon_{0}\pm g\mu_{B}B^c/2$ in this section. The time-dependent periodic component of the magnetic field $\vec{B}'(t)$ 
then creates additional states, i.e., sidebands, at energies $\epsilon_\uparrow\pm\omega$ and $\epsilon_\downarrow\pm\omega$ (see Fig. \ref{fig: energy levels}). 
These Zeeman levels and sidebands contribute to the elastic transport properties of the junction when their energies lie inside the 
bias-voltage window of $eV=\mu_{L}-\mu_{R}$. 

However, energy levels outside the bias-voltage window may also contribute to the electronic transport due to inelastic tunnel processes 
generated by the time-dependent magnetic field. In these inelastic processes, an electron transmitted from the left lead to the QD can 
change its energy by $\omega$ and either tunnel back to the left lead or out into the right lead. If this perturbation is small, as is 
assumed in this paper where we consider first-order corrections, the transport properties are still dominated by the elastic, energy-conserving 
tunnel processes that are associated with the Zeeman levels.

\begin{figure*}[t]
\includegraphics[height=5.62cm,keepaspectratio=true]{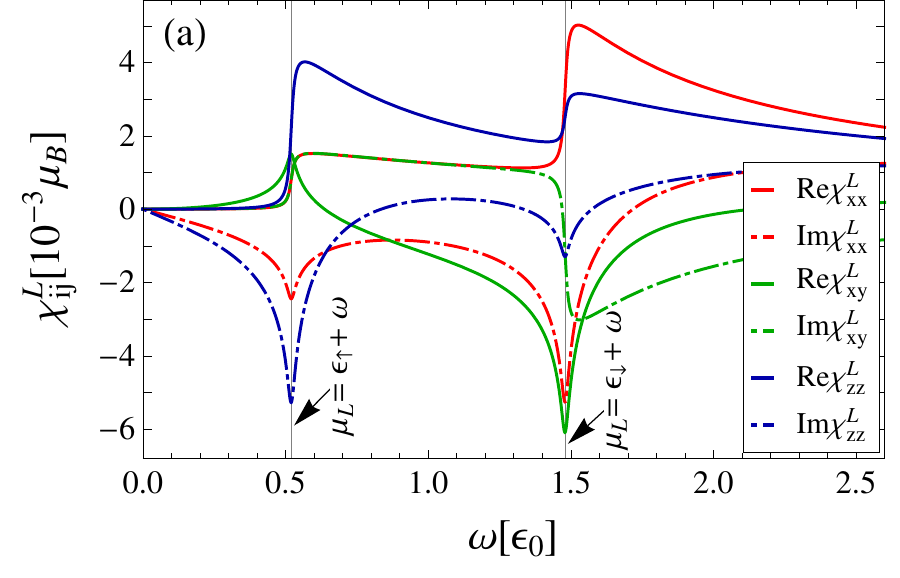}\,\,\,
\includegraphics[height=5.61cm,keepaspectratio=true]{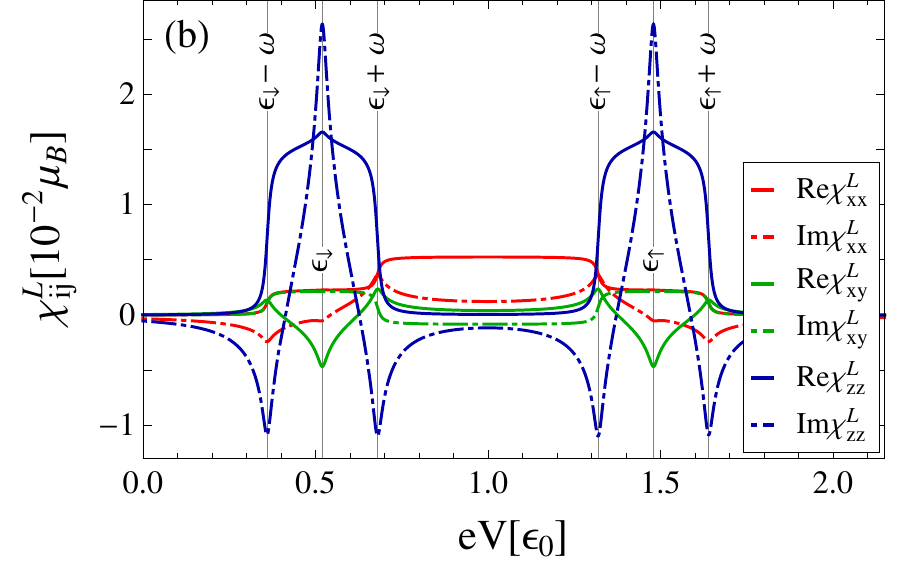}\,\,\,\,
\includegraphics[height=5.60cm,keepaspectratio=true]{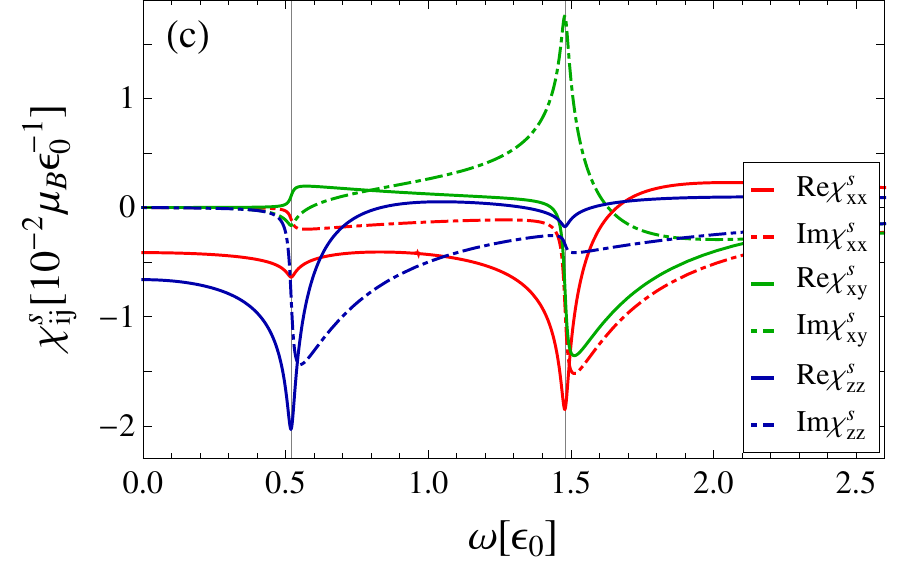}\,\,\,\,  
\includegraphics[height=5.61cm,keepaspectratio=true]{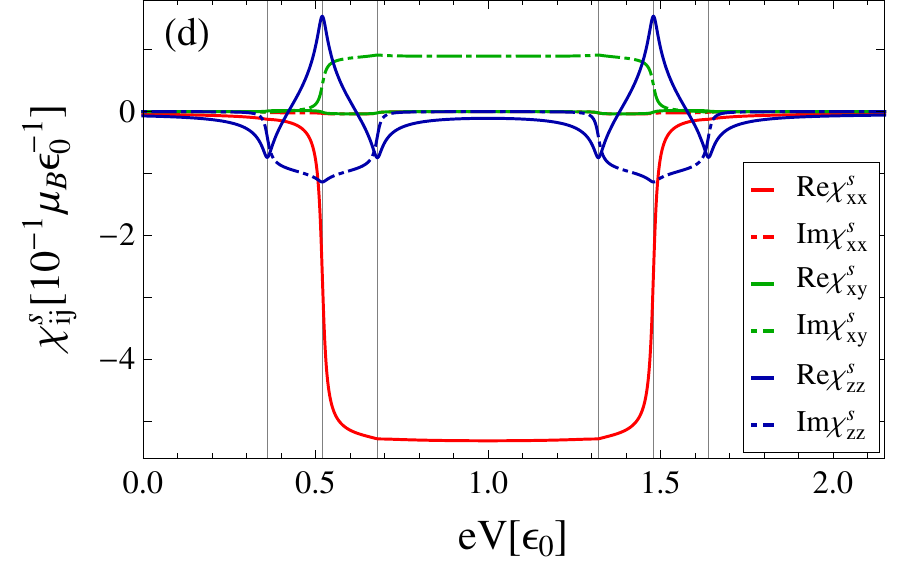}    
\caption{(Color online) (a) Frequency and (b) bias-voltage dependence of the spin-current susceptibilities. (c) Frequency and (d) bias-voltage 
dependence of the spin susceptibilities. In (a) and (c), the chemical potential of the left lead is $\mu_L=2\epsilon_0$, while in (b) and (d) 
the frequency is set to $\omega=0.16\epsilon_0$. All plots are obtained at zero temperature with $\vec B^c=B^c\vec e_{z}$, and the other 
parameters set to $\mu_{R}=0, \, \epsilon_{\uparrow}=1.48\epsilon_{0},\, \epsilon_{\downarrow}=0.52\epsilon_{0}, \, \Gamma=0.02\epsilon_{0}$, and $\Gamma_{L}=\Gamma_{R}=0.01\epsilon_{0}$.}\label{fig: susc}
\end{figure*}

The energy levels of the QD determine transport properties such as the spin-current susceptibilities and the spin susceptibilities, which are 
shown in Fig.~\ref{fig: susc}.
The imaginary and real parts of the susceptibilities are plotted as functions of the frequency $\omega$ in Figs. \ref{fig: susc}(a) and \ref{fig: susc}(c). 
In this case, the position of the unperturbed level $\epsilon_{0}$ is symmetric with respect to the Fermi surfaces of the leads and 
a peak or step in the spin-current and spin susceptibilities appears at a value of $\omega$, for which an energy level is aligned with one of 
the lead Fermi surfaces. In Figs. \ref{fig: susc}(b) and \ref{fig: susc}(d), the susceptibilities are instead plotted as functions of the bias voltage, $eV$.
Here, each peak or step in the susceptibilities corresponds to a change in the number of available transport channels.
The bias voltage is applied in such a way that the energy of the Fermi surface of the right lead is fixed at $\mu_{R}=0$ while the energy of the 
left lead's Fermi surface is varied according to $\mu_{L}=eV$.

\section{Spin-transfer torque and molecular spin dynamics}
\subsection{Model with a precessing molecular spin}

\begin{figure}[t]
\includegraphics[width=8.6cm,keepaspectratio=true]{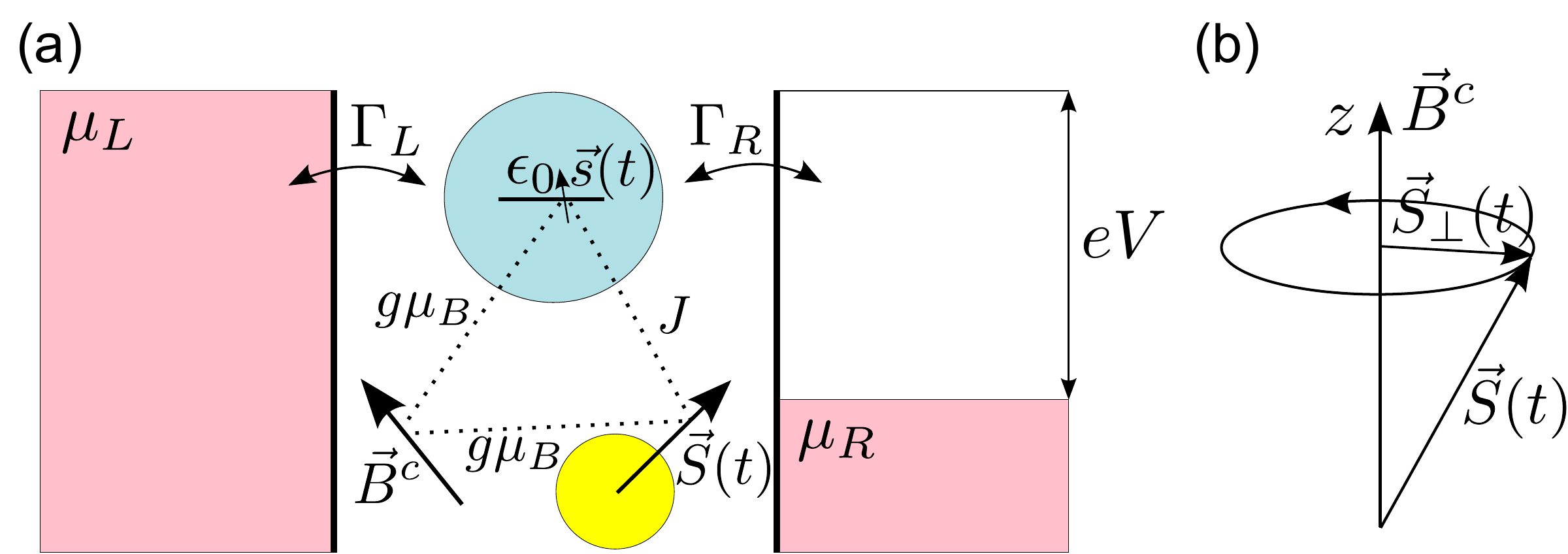}
\caption{(Color online) (a) Resonant tunneling in the presence of an SMM and an external, constant magnetic field. The electronic 
level of Fig. \ref{fig: qd and b field} is now coupled with the spin of an SMM via exchange interaction with the coupling constant $J$. 
The dynamics of the SMM's spin $\vec S$ is controlled by the external magnetic field $\vec B^c$ that also affects the electronic level.
(b) Precessional motion of the SMM's spin in a constant magnetic field $\vec B^c$ applied along the $z$ axis.}\label{fig: junction with SMM}
\end{figure}

Now we apply the formalism of the previous section to the case of resonant tunneling through a QD in the presence of 
a constant external magnetic field and an SMM [see Fig. \ref{fig: junction with SMM}(a)].
An SMM with a spin $S$ lives in a $(2S+1)$-dimensional Hilbert 
space. We assume that the spin $S$ of the SMM is large and 
neglecting the quantum fluctuations, one can
treat it as a classical vector whose end point moves on a sphere of radius $S$.
In the presence of a constant magnetic field ${\vec B}^c=B^c\vec e_{z}$,
the molecular spin precesses around the field axis
according to
$\vec S(t)=S_{\bot}\cos (\omega_{L} t)\vec e_x+S_{\bot}\sin (\omega_{L} t)\vec e_y+S_{z}\vec e_z$, where $S_{\bot}$ is the projection 
of $\vec S$ onto the $xy$ plane, $\omega_L=g\mu_{B}B^c $ is the Larmor precession frequency and $S_z$ is the projection of the 
spin on the $z$ axis [see Fig. \ref{fig: junction with SMM}(b)].
The spins of the electrons in the electronic level are coupled to the spin of the SMM via the exchange interaction $J$.
The contribution of the external magnetic field and 
the precessional motion of the SMM's spin create an effective time-dependent magnetic field acting on the electronic level.

The Hamiltonian of the system is now
given by $\hat{H}(t)=\hat{H}_{L,R}+\hat{H}_{T}+\hat{H}_D(t)+\hat{H}_{S}$, where the Hamiltonians $\hat{H}_{L,R}$ and $\hat{H}_{T}$ are 
the same as in Sec. \ref{sec: model}.
The Hamiltonian $\hat{H}_{S}=g\mu_{B}{\vec S} {\vec B}^c $ represents the interaction of the 
molecular spin $\vec S$ with the magnetic field $\vec B^c$ and consequently does not affect the electronic transport through the junction
but instead contributes to the spin dynamics of the SMM.
The Hamiltonian of the QD in this case is given by $\hat{H}_D(t)=\hat{H}^c_D+\hat{H}'(t)$. 
Here, $\hat{H}^c_D=\sum_{{\sigma}}\epsilon_{0}
\hat{d}^\dagger_{\sigma} \hat{d}_{\sigma}+g\mu_B\hat{\vec{s}} \vec{B}_{\rm eff}^c $ is the
Hamiltonian of the electrons in the QD in the presence of 
the constant part of the effective magnetic field, given by $\vec B_{\rm eff}^c =\big [B^c+\frac{J}{g\mu_B}S_{z}\big ]\vec e_{z}$.
The second term of the QD Hamiltonian,
$\hat{H}'(t)=g\mu_{B}\hat{\vec{s}} \vec{B}'_{{\rm eff}}(t)$, represents the interaction between the electronic spins of the QD, $\hat{\vec{s}}$, 
and the time-dependent part of the effective magnetic field, given by 
$\vec B'_{{\rm eff}}(t)=\frac{JS_{\bot}}{g\mu_{B}}\big [\cos (\omega_{L} t)\vec e_{x}+\sin (\omega_{L} t)\vec e_{y}\big ]$. The time-dependent
effective magnetic field can be rewritten as $\vec{B}'_{{\rm eff}}(t)= \vec B_{\omega_L}e^{-i\omega_{L} t}+\vec B_{\omega_L}^{\ast}e^{i\omega_{L} t}$, 
where $\vec B_{\omega_{L}}$ 
consists of the complex amplitudes $B_{\omega_{L} x}=JS_{\bot}/2g\mu_{B}$, 
$B_{\omega_{L} y}=iJS_{\bot}/2g\mu_{B}$, and $B_{\omega_{L} z}=0$.
The time-dependent perturbation can then be expressed as $\hat{H}^\prime(t)={\hat{H}}_{\omega_L} e^{-i\omega_{L} t}+{\hat{H}}^{\dagger}_{\omega_L} e^{i\omega_{L} t}$, 
where ${\hat{H}}_{\omega_L}$ is an operator that can be written, using Eq. (3) and the above expressions for $B_{\omega_{L} i}$, as
\begin{equation}\label{eq: H op}
{\hat{H}}_{\omega_L}=\frac{JS_{\bot}}{2}\left (
   \begin{array}{cc}
   0 & 1 \\
  0 & 0
   \end{array}
   \right ).
\end{equation}

The time-dependent part of the effective magnetic field creates inelastic tunnel processes that contribute to the currents.
The in-plane components of the spin current fulfill
\begin{align}\label{eq: xy spin currents}
I_{Lx}(\omega_L)&=-iI_{Ly}(\omega_L)\\
&=\frac{JS_{\bot}}{2g\mu_{B}}[\chi_{xx}^L(\omega_L)+i\chi_{xy}^L(\omega_L)],\nonumber
\end{align}
where $\vec{B}^{c}$ is replaced by $\vec{B}^{c}_{\rm eff}$.
The $z$ component vanishes to lowest order in $H'(t)$.\cite{W} 
Therefore, the inelastic spin current has a polarization that precesses in the $xy$ plane.
The inelastic spin-current components, in turn, exert an STT (Refs. 41-44) 
on the molecular spin given by
 \begin{equation}\label{eq: torque}
\vec T(t)= -[{\vec I}_L(t)+{\vec I}_R(t)],
 \end{equation}
thus contributing to the dynamics of the molecular spin through
\begin{eqnarray}
\dot{\vec{S}}(t)=g\mu_{B}\vec B^c\times\vec{S}(t)+\vec T(t).
\end{eqnarray}

Using expressions (\ref{eq: J of t}), (\ref{eq: spin current}), and (\ref{eq: H op}), the torque of Eq. (\ref{eq: torque}) can be calculated in 
terms of the Green's functions $\hat{G}_{0}^{r}(\epsilon)$ and $\hat{G}_{0}^{a}(\epsilon)$ as
\begin{align}\label{eq: torque T}
T_{i}(t)=&-\frac{JS_{\bot}}{2}\int\frac{d\epsilon}{2\pi}\sum_{\xi\lambda}\frac{\Gamma_{\xi}
\Gamma_{\lambda}}{\Gamma}[f_{\xi}(\epsilon-\omega_L)-f_{\lambda}(\epsilon)]\\
&\times{\rm Im}\{({{\hat\sigma}}_{i})_{\downarrow\uparrow}G^{r}_{0,\uparrow\uparrow}(\epsilon)G^{a}_{0,\downarrow\downarrow}(\epsilon-\omega_L)e^{-i\omega_{L} t}\},\nonumber
\end{align} 
with $\lambda=L,R$.
Here $(\hat\sigma_{i})_{\downarrow\uparrow}$, $G^{r}_{0,\uparrow\uparrow}(\epsilon)$, and $G^{a}_{0,\downarrow\downarrow}(\epsilon)$ are matrix elements
of ${\hat\sigma}_{i}$, ${\hat{G}}_{0}^{r}(\epsilon)$ and ${\hat{G}}_{0}^{a}(\epsilon)$ with respect
to the basis of eigenstates of ${\hat{s}}_z$.
This STT can be rewritten in terms of the SMM's spin vector as
\begin{eqnarray}
\vec T(t)=\frac{\alpha}{S}\dot{\vec{S}}(t)\times\vec{S}(t)+\beta\dot{\vec{S}}(t)+\gamma\vec{S}(t).
\end{eqnarray}
The first term in this back-action 
gives a contribution to the Gilbert damping, characterized by the Gilbert 
damping coefficient $\alpha$.
The second term acts as an effective constant magnetic field and changes the precession frequency of the spin $\vec S$ with the corresponding coefficient $\beta$.
The third term cancels the $z$ component of the Gilbert damping term, thus restricting the 
STT to the $xy$ plane. The coefficient of the third term $\gamma$ is related
to $\alpha$ by $\gamma/\alpha=\omega_{L} S^2_\perp/SS_z$. 
Expressing the coefficients $\alpha$ and $\beta$ in terms of the current susceptibilities $\chi^{\xi}_{xx}(\omega_L)$ and  $\chi^{\xi}_{xy}(\omega_L)$ results in
\begin{align}
\alpha=&-\frac{JS_{z}}{g\mu_{B}\omega_{L} S}\sum_{\xi}[{\rm Re}\{\chi^{\xi}_{xx}(\omega_L)\}-{\rm Im}\{\chi^{\xi}_{xy}(\omega_L)\}], \\
\beta=&\frac{J}{g\mu_{B}\omega_L}\sum_{\xi}[{\rm Im}\{\chi^{\xi}_{xx}(\omega_L)\}+{\rm Re}\{\chi^{\xi}_{xy}(\omega_L)\}].\ 
\end{align}
By inserting the explicit expressions for $G^{r}_{0,\uparrow\uparrow}(\epsilon)$ and $G^{a}_{0,\downarrow\downarrow}(\epsilon-\omega_L)$, one obtains
\begin{align}\label{eq: alpha explicit}
\alpha=&\frac{J^{2}S_z^{2}}{\omega_{L} S}\int\frac{d\epsilon}{8\pi}\sum_{\xi\lambda}
\Gamma_{\xi}\Gamma_{\lambda}[f_{\xi}(\epsilon-\omega_L)-f_{\lambda}(\epsilon)]\times\\
&\frac{1}{[(\frac{\Gamma}{2})^{2}+(\epsilon-\epsilon_{\uparrow})^{2}][(\frac{\Gamma}{2})^{2}+(\epsilon-\epsilon_{\downarrow}-\omega_L)^{2}]}, \nonumber
\end{align} 
\begin{align}
\beta=&-\frac{J}{\omega_{L}\Gamma}\int\frac{d\epsilon}{4\pi}\sum_{\xi\lambda}\Gamma_{\xi}\Gamma_{\lambda}[f_{\xi}(\epsilon-\omega_L)-f_{\lambda}(\epsilon)]\times\\
&\frac{(\frac{\Gamma}{2})^{2}+(\epsilon-\epsilon_{\uparrow})(\epsilon-\epsilon_{\downarrow}-\omega_L)}
{[(\frac{\Gamma}{2})^{2}+(\epsilon-\epsilon_{\uparrow})^{2}][(\frac{\Gamma}{2})^{2}+(\epsilon-\epsilon_{\downarrow}-\omega_L)^{2}]},\nonumber
\end{align}
where $\epsilon_{\uparrow,\downarrow}=\epsilon_{0}\pm g\mu_{B}B_{\rm eff}^c/2=\epsilon_{0}\pm (\omega_{L}+JS_{z})/2$ are the energies of the Zeeman levels
in this section.
In the small precession frequency regime, $\omega_{L}\ll k_BT$, $\gamma\rightarrow 0$ and in the limit of $S_{z}/S \rightarrow 1$ 
the expression for the coefficient $\alpha$ 
is in agreement with Ref. \onlinecite{Bode}.

\subsection{Analysis of the spin-transfer torque}

\begin{figure}[t]
\includegraphics[height=4.2cm,keepaspectratio=true]{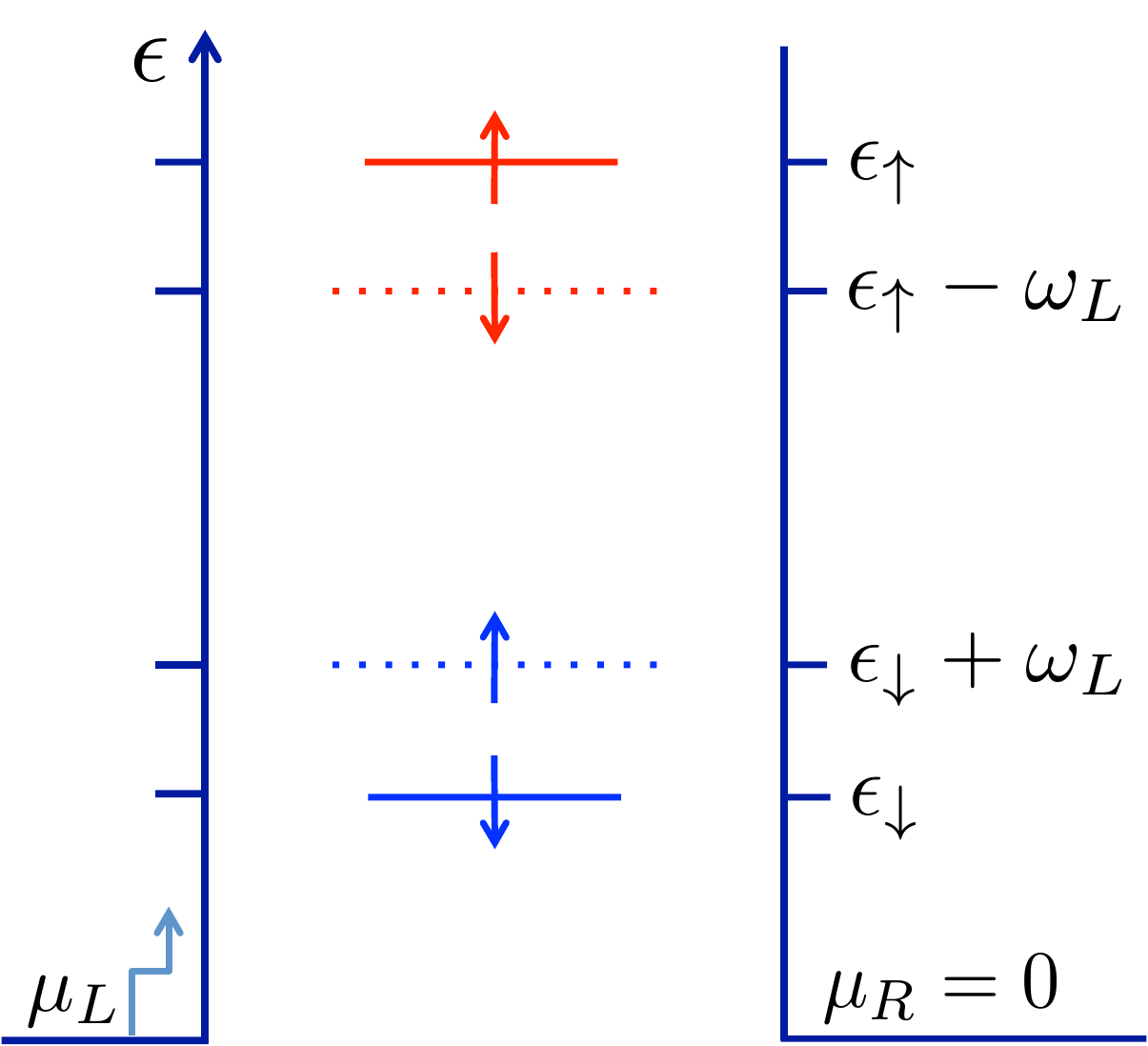}
\caption{(Color online) Sketch of the electronic energy levels of the QD in the
presence of a molecular spin precessing with the frequency $\omega_L$ around an external, constant magnetic field.
The corresponding Zeeman levels are $\epsilon_{\uparrow,\downarrow}$.
The precessional motion of the molecular spin results in emission (absorption) of 
energy corresponding to a spin flip from spin up (down) to spin down (up). Hence, there are only four channels
available for transport.}\label{fig: energy levels1}
\end{figure}

\begin{figure}[t]
\includegraphics[height=6.0cm,keepaspectratio=true]{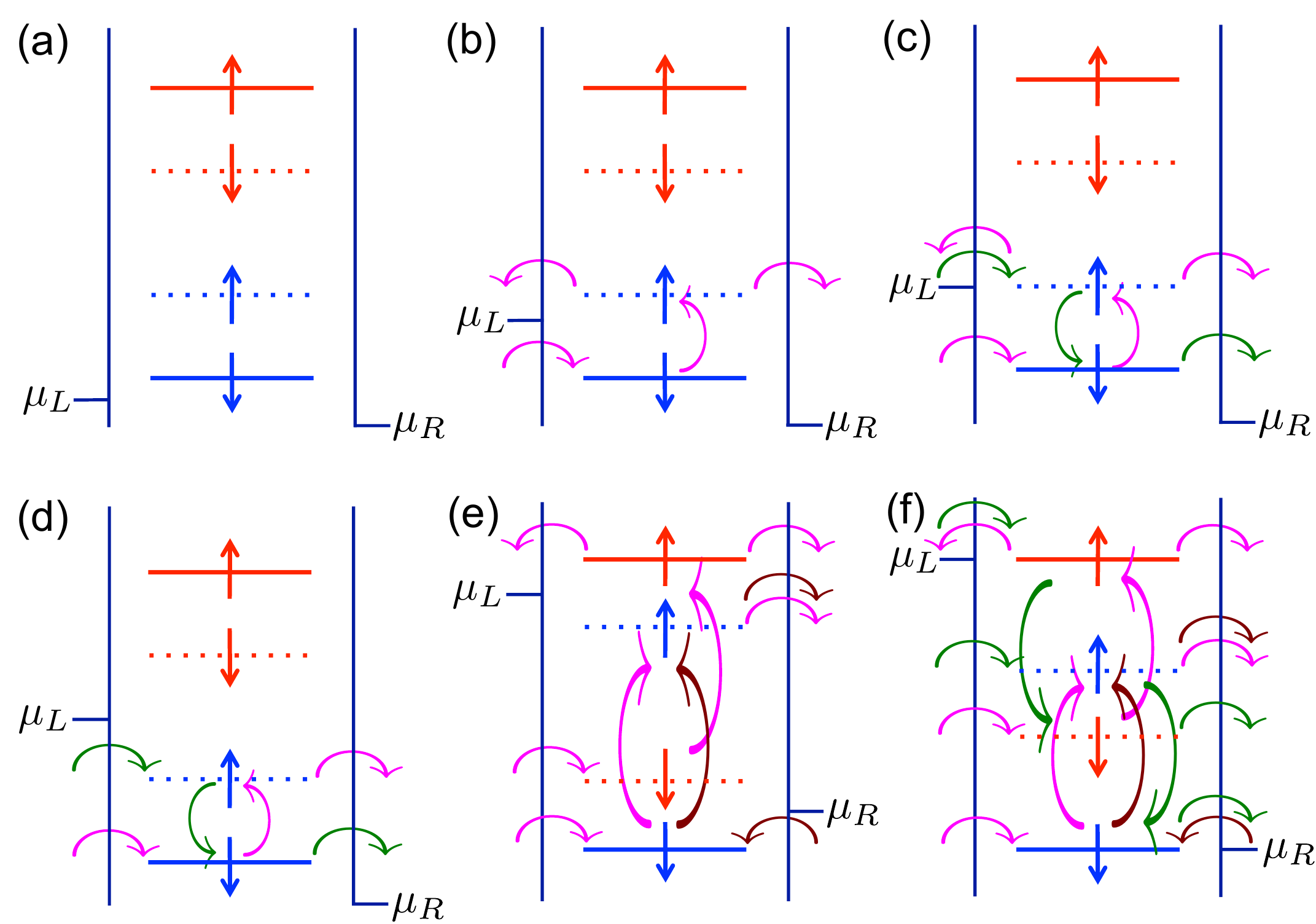}
\caption{(Color online) Sketch of the inelastic spin-tunneling processes in the QD in
the presence of the precessing molecular spin in the field $\vec B^c=B^c\vec e_{z}$ for different
positions of the energy levels with respect to the chemical potentials of the leads, $\mu_{L}$ and $\mu_{R}$.
Only transitions between levels with the same color (blue or red) are allowed. Different colored
curved arrows (magenta, brown, or green) represent different processes.}\label{fig: inelastic processes}
\end{figure}

In the case of resonant tunneling in the presence of a molecular spin precessing in a constant external magnetic field, one also needs to take 
the exchange of spin-angular momentum between the molecular spin and the electronic spins into account in addition to the effects of the 
external magnetic field. 
Due to the precessional motion of the 
molecular spin, an electron in the QD emitting (absorbing) an energy $\omega_L$ also undergoes a spin flip from spin up (down) to spin down (up), 
as indicated by the arrows in Fig. \ref{fig: energy levels1}. As a result, the levels at 
energies $\epsilon_{\uparrow,\downarrow}\pm\omega_L$ are 
forbidden and hence do not contribute to the transport processes. Consequently, there are only four transport channels, which are located  
at energies $\epsilon_{\uparrow,\downarrow}\mp\omega_L$.
Also in this case, there are elastic and inelastic tunnel processes.
Some of the possible inelastic tunnel processes are shown in Fig. \ref{fig: inelastic processes}.
These restrictions on the inelastic tunnel processes are also visible in Fig. \ref{fig: susc}(b),
which identically corresponds to the case of the presence of a precessing molecular spin with $\omega_{L}=0.16\epsilon_{0}$ and $JS_{z}=0.8\epsilon_{0}$. 
Namely, from Eq. (\ref{eq: xy spin currents}), which is equivalent to
${\rm Re}\{ I_{Lx}(\omega_L)\}={\rm Im} \{I_{Ly}(\omega_L)\}=\frac{JS_{\bot}}{2g\mu_{B}}[{\rm Re}\{\chi_{xx}^L(\omega_L)\}-{\rm Im}\{\chi_{xy}^L(\omega_L)\}]$ and 
${\rm Im} \{I_{Lx}(\omega_L)\}=-{\rm Re} \{I_{Ly}(\omega_L)\}=\frac{JS_{\bot}}{2g\mu_{B}}[{\rm Im}\{\chi_{xx}^L(\omega_L)\}+{\rm Re}\{\chi_{xy}(\omega_L)\}]$,
and from the symmetries of the susceptibilities displayed in Fig. \ref{fig: susc}(b), it follows that there are no 
spin currents at $eV=\epsilon_{\uparrow,\downarrow}\pm\omega_L$.

\begin{figure*}[t]
\includegraphics[height=6.0cm,keepaspectratio=true]{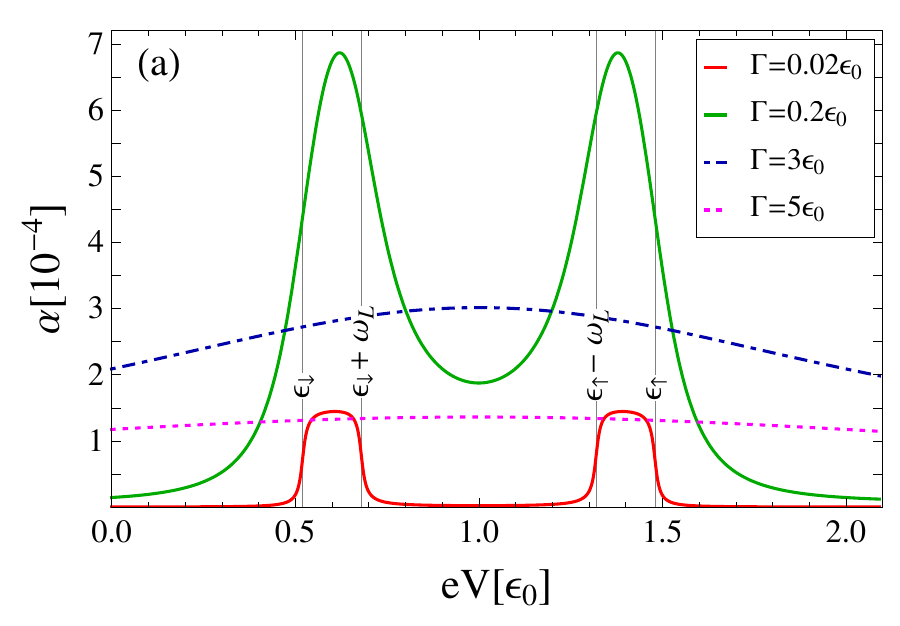}\,\,\,\,
\includegraphics[height=6.0cm,keepaspectratio=true]{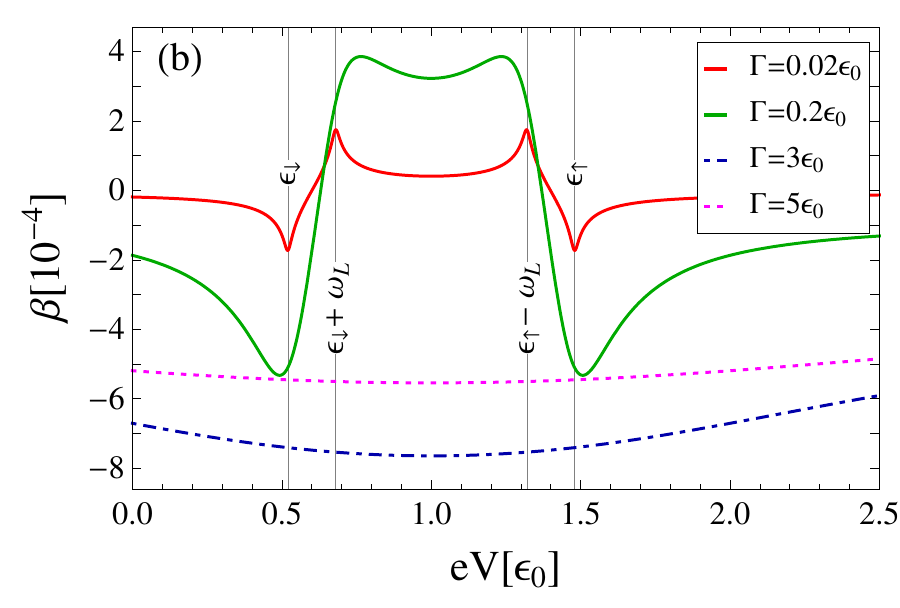}
\caption{(Color online) (a) Gilbert damping coefficient $\alpha$ and (b) coefficient $\beta$ as functions of the applied bias 
voltage $eV=\mu_{L}-\mu_{R}$, with $\mu_{R}=0$, for different tunneling rates $\Gamma$ at zero temperature. Other 
parameters are $\Gamma_{L}=\Gamma_{R}=\Gamma/2$, $\epsilon_{\uparrow}=1.48\epsilon_{0}$, $\epsilon_{\downarrow}=0.52\epsilon_{0}$, $S=100$, $J=0.01\epsilon_{0}$, $JS_{z}=0.8\epsilon_{0}$, and $\omega_L=0.16\epsilon_{0}$. 
In the case of the smallest value of $\Gamma$ (red lines), $\alpha$ approaches a constant value when $\mu_L$ lies within the energy range specified by Eqs. (25) and (26).
The coefficient $\beta$ has one local minimum and one local maximum for the same energy range.}\label{fig: gilbert and beta}
\end{figure*}
\begin{figure*}[t]
\includegraphics[height=5.95cm,keepaspectratio=true]{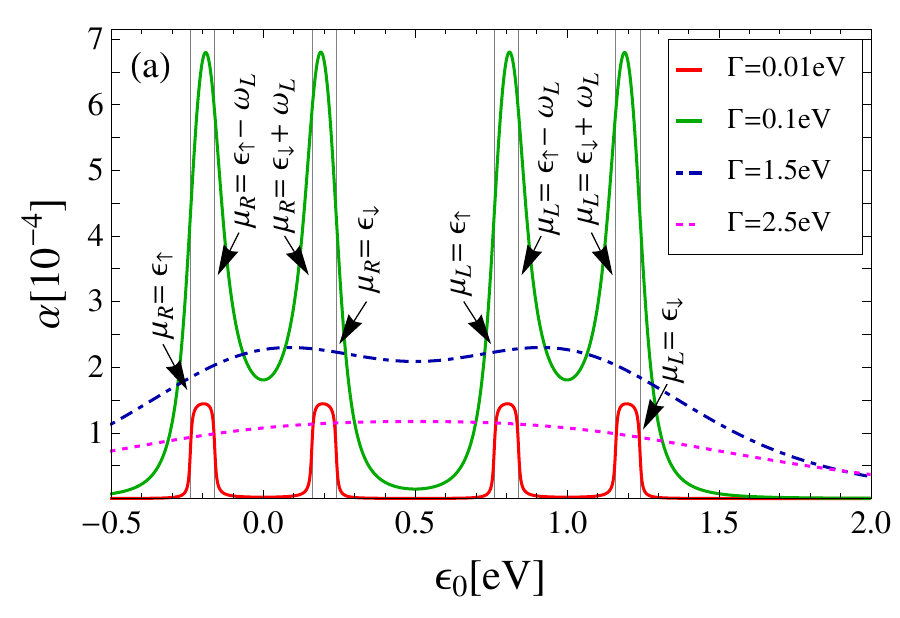}\,\,\,\,
\includegraphics[height=5.95cm,keepaspectratio=true]{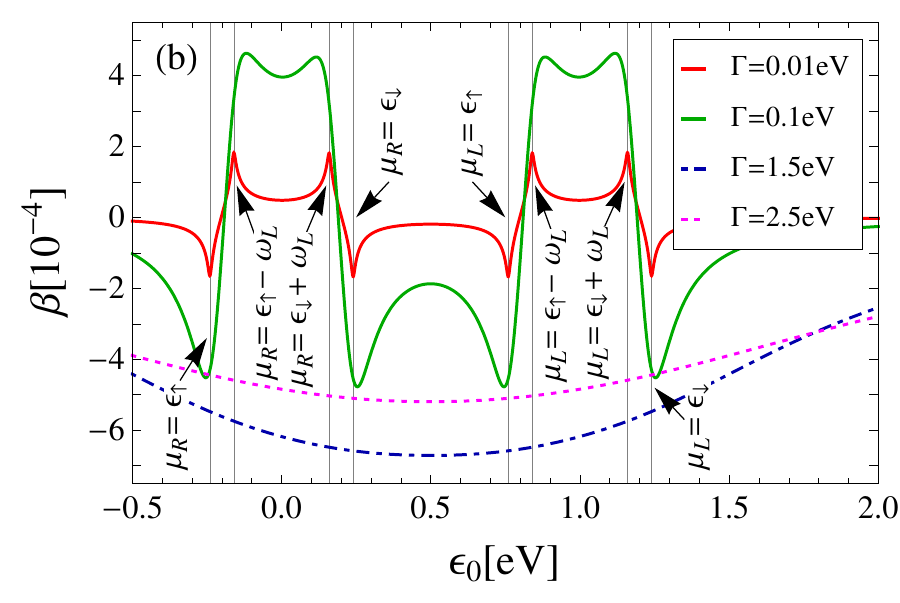}
\caption{(Color online) (a) Gilbert damping coefficient $\alpha$ and (b) coefficient $\beta$ as functions of the position of the electronic level $\epsilon_{0}$ for 
different tunneling rates $\Gamma$ at zero temperature. The applied bias voltage 
is $eV=\mu_{L}-\mu_{R}$, with $\mu_{R}=0$. Other parameters 
are $\Gamma_{L}=\Gamma_{R}=\Gamma/2$, $\epsilon_{\uparrow}-\epsilon_{0}=0.24 eV$, $S=100$, $J=0.005 eV$, $JS_{z}=0.4 eV$, and $\omega_L=0.08 eV$. In the case of 
the smallest value of $\Gamma$ (red lines), there are four regions in which the Gilbert damping and the change of the precession frequency 
occur. In each of these regions $\epsilon_{0}$ satisfies the inequalities (25) and (26), and $\alpha$ approaches a constant value, 
while $\beta$ has one local maximum and one local minimum.}\label{fig: gilbert and beta 2}
\end{figure*}

As was mentioned, the spin currents generate an STT acting on the molecular spin. A necessary condition for the existence of an STT, 
and hence finite values of the coefficients $\alpha$ and $\beta$ in Eqs. (23) and (24), is 
that $\vec I_L (t)\neq -\vec I_R(t)$ [see Eq. (\ref{eq: torque})]. This condition is met by the spin currents generated, e.g., by the 
inelastic tunnel processes shown in Figs. \ref{fig: inelastic processes}(b) and \ref{fig: inelastic processes}(c). These tunnel processes occur when an electron can 
tunnel into the QD, undergo a spin flip, and then tunnel off the QD into either lead. From these tunnel processes it is implied that the 
Gilbert damping coefficient $\alpha$ and the coefficient $\beta$ can be controlled by the applied bias or gate voltage as well as by the external magnetic field.
If a pair of QD energy levels, coupled via spin-flip processes, lie within the bias-voltage window, 
the spin currents instead fulfill
$\vec I_L(t)=\vec I_R(t)$, leading to a vanishing STT [see Fig. \ref{fig: inelastic processes}(d)].
In Figs. \ref{fig: inelastic processes}(e) and \ref{fig: inelastic processes}(f) the position of the energy levels of the QD are symmetric with 
respect to the Fermi levels
of the leads, $\mu_{L}$ and $\mu_{R}$.
When the QD level with energy $\epsilon_\uparrow$ is aligned 
with $\mu_L$, this simultaneously corresponds to the energy level $\epsilon_\downarrow$ being aligned with $\mu_R$  
[see Fig. \ref{fig: inelastic processes}(f)]. As a result, a spin-up electron 
can now tunnel 
from the left lead into the level $\epsilon_{\uparrow}$, 
while a spin-down electron in the level $\epsilon_\downarrow$ can tunnel into
the right lead. These additional processes enhance the STT compared to that of the case \ref{fig: inelastic processes}(e).

\begin{figure*}[t]
\includegraphics[height=6.0cm,keepaspectratio=true]{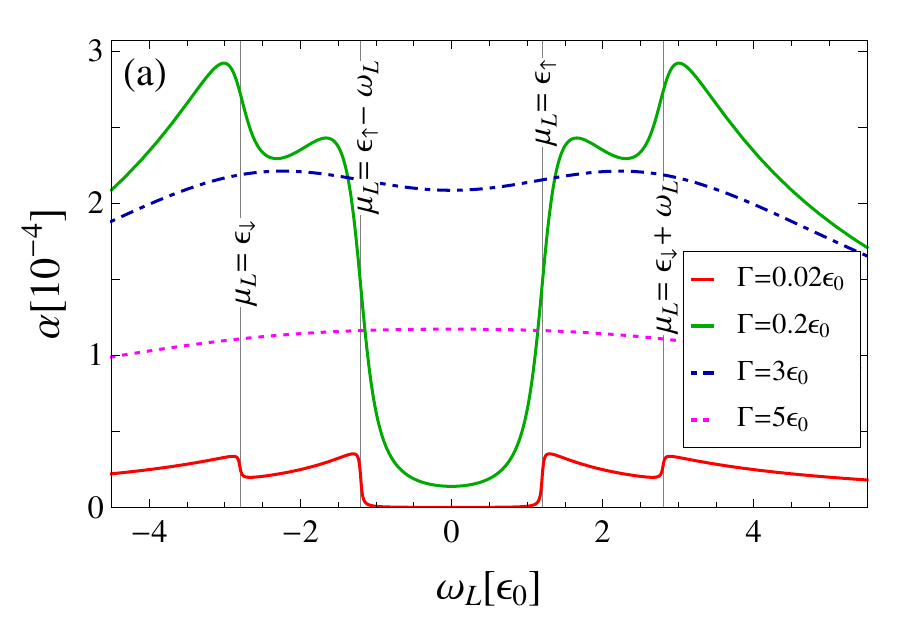}\,\,\,\,
\includegraphics[height=6.0cm,keepaspectratio=true]{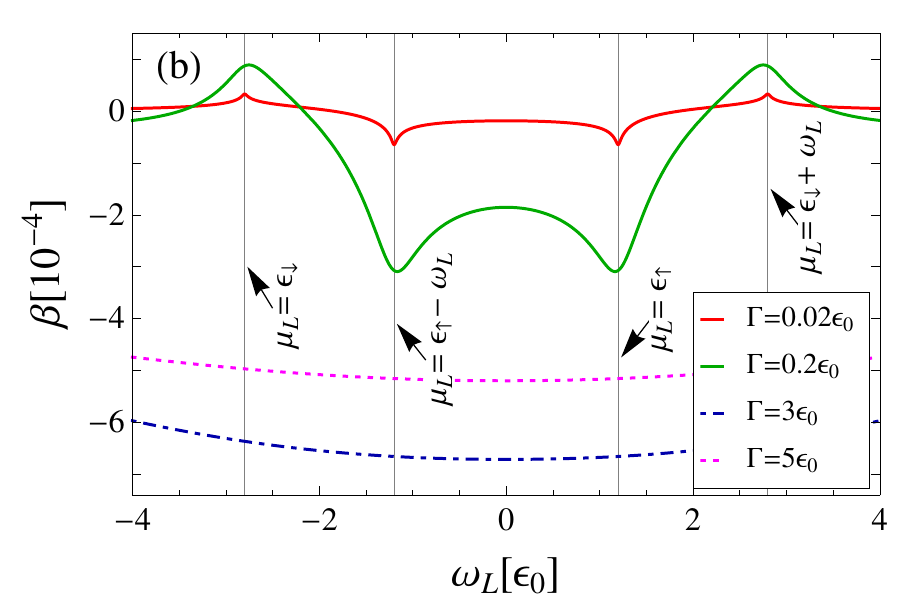}
\caption{(Color online) (a) Gilbert damping coefficient $\alpha$ and (b) coefficient $\beta$ as functions of the precession 
frequency $\omega_L=g\mu_{B} B^c $ of the spin $\vec S$ of the SMM, with $\vec B^c=B^c\vec e_{z}$, for different tunneling rates $\Gamma$ at zero temperature. 
The applied bias voltage is $eV=\mu_{L}-\mu_{R}=2\epsilon_{0}$, with $\mu_{R}=0$. The other parameters are the same as in Fig. 7. In the case of the 
smallest $\Gamma$ (red lines), the  coefficient $\alpha$ has a step increase towards a local maximum while the coefficient $\beta$ has a local maximum or minimum at a value of $\omega_{L}$ corresponding to a resonance of $\mu_{L}$ with one of the levels in the QD.}\label{fig: gilbert and beta 3}
\end{figure*}

The two spin-torque coefficients $\alpha$ and $\beta$ exhibit a nonmonotonic dependence on the tunneling rates $\Gamma$, as can be seen in 
Figs. \ref{fig: gilbert and beta}, \ref{fig: gilbert and beta 2}, and \ref{fig: gilbert and beta 3}.
For $\Gamma\rightarrow 0$, it is obvious that $\alpha,\beta\rightarrow0$.
In the weak coupling limit $\Gamma\ll\omega_L$, 
the coefficients $\alpha$ and $\beta$ are finite if the Fermi surface energy of the lead $\xi$, $\mu_\xi$ fulfills either of the conditions
\begin{align}
\epsilon_{\downarrow}&\leq\mu_\xi\leq\epsilon_{\downarrow}+\omega_L   \label{eq: ineq 1} \\
{\rm or}\quad\epsilon_{\uparrow}-\omega_L&\leq\mu_\xi\leq\epsilon_{\uparrow}   \label{eq: ineq 2} \
\end{align} 
in such a way that each condition is satisfied by the Fermi energy of maximum one lead.
These conditions are relaxed for larger tunnel couplings as a consequence of the broadening of the QD energy levels, which is also
responsible for the initial enhancement of $\alpha$ and $\beta$ with increasing $\Gamma$. Notice, however, that $\alpha$ and $\beta$
are eventually suppressed for $\Gamma\gg\omega_L$, when
the QD energy levels are significantly broadened and overlap so that spin-flip processes 
are equally probable in each direction and there is no net effect on the molecular spin. 
Physically, this suppression of the STT can be understood by noticing that for $\Gamma\gg\omega_L$ a current-carrying electron 
perceives the molecular spin as almost static due to its slow precession compared to
the electronic tunneling rates and hence the exchange of angular momenta is reduced. 
With increasing tunneling rates, the coefficient $\beta$ becomes negative before it drops to zero, causing the torque $\beta\dot{\vec{S}}$ to oppose the 
rotational motion of the spin $\vec S$.

In Fig. \ref{fig: gilbert and beta}, the Gilbert damping coefficient $\alpha$ and the coefficient $\beta$ are plotted as  
functions of the applied bias voltage at zero temperature.
We analyze the case of the smallest value of $\Gamma$ (red lines), assuming that $\omega_{L}>0$. 
For small $eV$, all QD energy levels lie outside the bias-voltage window and there is no spin transport [see Fig. 6(a)]. Hence $\alpha,\beta\rightarrow 0$.
At $eV=\epsilon_\downarrow$ the tunnel processes in Fig. 6(b) come into play, leading to a finite STT and the coefficient $\alpha$ increases while the
coefficient $\beta$ has a local minimum. In the voltage region specified by Eq. (25) for $\mu_{L}$, the coefficient $\alpha$ approaches a constant value
while the coefficient $\beta$ increases. By increasing the bias voltage to $eV=\epsilon_\downarrow+\omega_L$ the tunnel processes in Fig. 6(c) occur, leading to a decrease of $\alpha$ and a local 
maximum of $\beta$. For $\epsilon_\downarrow+\omega_L<eV<\epsilon_\uparrow-\omega_L$, the coefficients $\alpha,\beta\rightarrow 0$ [see Fig 6(d)]. In the voltage region specified by Eq. (26) for
$\mu_{L}$, $\alpha$ approaches the same constant value mentioned above while $\beta$ decreases between a local maximum at $eV=\epsilon_\uparrow-\omega_L$ and a
local minimum at $eV=\epsilon_\uparrow$, which approach the same values as previously mentioned extrema. With further increase of $eV$, all QD energy levels lie within the bias-voltage window
and the STT consequently vanishes. 

Figure \ref{fig: gilbert and beta 2} shows the spin-torque coefficients $\alpha$ and $\beta$
as  functions of the position of the electronic level $\epsilon_{0}$. 
An STT acting on the molecular spin occurs if the electronic level $\epsilon_{0}$ is positioned in such a way that the inequalities (\ref{eq: ineq 1}) 
and (\ref{eq: ineq 2}) may be satisfied by some values of $eV$, $\epsilon_0$ and $\omega_L$.
Again, we analyze the case of the smallest value of $\Gamma$ (red curve).
For the particular choice of parameters in Fig. \ref{fig: gilbert and beta 2}, there are four regions in which the inequalities (\ref{eq: ineq 1}) 
and (\ref{eq: ineq 2})
are satisfied. Within these regions, $\alpha$ approaches a constant value while $\beta$ has a local maximum as well as a local minimum. These local extrema 
occur when one of the Fermi surfaces is aligned with one of the energy levels of the QD.
For other values of $\epsilon_{0}$, both $\alpha$ and $\beta$ vanish.

The coefficients $\alpha$ and $\beta$ are plotted as functions of the precession frequency $\omega_L$ in Fig. \ref{fig: gilbert and beta 3}.
Here, $\epsilon_{0}=eV/2$ and therefore the positions of the energy levels of the
QD are symmetric with respect to the Fermi levels of the leads, $\mu_{L}$ and $\mu_{R}$.
Once more, we focus first on the case of the smallest value of $\Gamma$ (indicated by the red curve).
The energies of all four levels of the QD depend on $\omega_L$, i.e., $\vec{B}^{c}$.
For $\omega_{L}>0$, when the magnitude of the external magnetic field is large enough, 
the tunnel processes in Fig. \ref{fig: inelastic processes}(f) take place due to the above-mentioned symmetries.
These tunnel processes lead to a finite STT, a maximum for 
the Gilbert damping coefficient $\alpha$, and a negative minimum value for the $\beta$ coefficient. 
As $\omega_L$ increases,
the inequalities of Eqs. (\ref{eq: ineq 1}) and (\ref{eq: ineq 2}) are satisfied and the tunnel processes shown in Fig. \ref{fig: inelastic processes}(e) 
may occur. Hence, there is a contribution to the STT, but as is shown in Eq. (\ref{eq: alpha explicit}), the Gilbert damping 
decreases with increasing precession frequency.
At larger values of $\omega_L$, resulting in ${\epsilon_\downarrow+\omega_L=\mu_L}$, the Gilbert damping coefficient has a step increase towards a local maximum, while the 
coefficient $\beta$ has a local maximum, as a consequence of the enhancement of the STT due to additional spin-flip processes occurring in this case. 
For even larger value of $\omega_L$, the conditions (\ref{eq: ineq 1}) and (\ref{eq: ineq 2}) are no 
longer fulfilled and both coefficients vanish.
It is energetically unfavorable to flip the spin of an electron against the antiparallel direction of the effective constant magnetic field $B_{\rm eff}^c $. 
Hence, as $\omega_L$ increases, more energy is needed to flip the electronic spin to the
direction of the field.
This causes $\alpha$ to decrease with increasing $\omega_L$.
Additionally, the larger the ratio $\omega_{L}/\Gamma$, the less probable it is that spin-angular momentum will be exchanged between the molecular spin and 
the itinerant electrons.
For $\omega_{L}=0$, the molecular spin is static, i.e., $\dot{\vec{S}}=0$. In this case $\vec{T}(t)=\vec{0}$.
The coefficient $\alpha$ then drops
to zero while the coefficient $\beta$ reaches a negative local maximum which is close to 0. 
Both $\alpha$ and $\beta$ reach an extremum value for large values of $\Gamma$ at this point.
For $\omega_{L}<0$ and $\Gamma\ll\lvert\omega_{L}\rvert$ (red lines), at 
the value of $\omega_L$ for which $\mu_{L}=\epsilon_\uparrow-\omega_L$, the coefficient $\alpha$ has a 
step increase towards a local maximum while the coefficient $\beta$ has a negative local minimum. The coefficient $\alpha$ 
then decreases with a further decrease of $\omega_L$ as long as 
$\epsilon_{\downarrow}\leq\mu_{L}\leq\epsilon_\uparrow-\omega_L$.
At the value of $\omega_L$ for which $\mu_{L}=\epsilon_\downarrow$, $\alpha$ 
has another step increase towards a local maximum
while $\beta$ has a maximum value.
According to Eq. (\ref{eq: alpha explicit}), the Gilbert damping also does not occur if $\vec{S}$ is 
perpendicular to $\vec B^c$. In this case $\beta\lesssim 0$ and the only nonzero torque component $\beta\dot{\vec{S}}(t)$ acts in the oposite direction than the molecular spin's rotational motion.

\section{Conclusions}

In this paper we have first theoretically studied time-dependent charge and spin transport through a small junction consisting of a single-level quantum dot 
coupled to two noninteracting metallic leads in the presence of a time-dependent magnetic field.
We used the Keldysh nonequilibrium Green's functions method to derive the charge and spin currents in linear order with respect to the 
time-dependent component of the magnetic field with a characteristic frequency $\omega$.
We then focused on the case of a single electronic level coupled via exchange interaction to an effective magnetic 
field created by the precessional motion of an SMM's spin in a constant magnetic field. 
The inelastic tunneling processes that contribute to the spin currents produce an STT
that acts on the molecular spin.
The STT consists of a Gilbert damping component, characterized by the coefficient $\alpha$, as well as a component, characterized by the coefficient $\beta$, that 
acts as an additional effective constant magnetic 
field and changes the precession frequency $\omega_L$ of the molecular spin.
Both $\alpha$ and $\beta$ depend on $\omega_L$ and show a nonmonotonic dependence on the tunneling rates $\Gamma$.
In the weak coupling limit $\Gamma\ll\omega_L$, $\alpha$  
can be switched on and off as a function of bias and gate voltages. 
The coefficient $\beta$ correspondingly has a local extremum.
For $\Gamma\rightarrow0$, both $\alpha$ and $\beta$ vanish.
Taking into account that spin transport can be controlled by the bias and gate voltages, as well as by external magnetic fields, 
our results might be useful in spintronic applications using SMMs.
Besides a spin-polarized STM, it may be possible to detect and manipulate the spin state of an SMM in a ferromagnetic resonance 
experiment\cite{Costache,Aarts,Harder,Ando} and thus extract information about the effects of the current-induced STT on the SMM.
Our study could be complemented with a quantum description of an SMM in a single-molecule magnet junction and its coherent properties, as these render
the SMM suitable for quantum information storage.

\begin{acknowledgments}
We gratefully acknowledge discussions with Mihajlo Vanevi\'{c} and Christian Wickles. 
This work was supported by Deutsche Forschungsgemeinschaft through SFB 767. We are thankful for
partial financial support by an ERC Advanced Grant, project \textit{UltraPhase} of Alfred Leitenstorfer.
\end{acknowledgments}


\begin{thebibliography}{12}
\bibitem{W}
However, $z$-polarized dc spin currents 
appear in second-order approximations. \cite{Guo}
\end{thebibliography}
\end{document}